%% file: cas-dc-template.tex
\documentclass[a4paper,fleqn]{cas-dc}

\usepackage[numbers]{natbib}
\usepackage{caption}
\usepackage{subcaption}
\usepackage{amsmath}
\usepackage{mathtools}
\usepackage[ruled,vlined]{algorithm2e}
\DeclarePairedDelimiterX{\infdivx}[2]{(}{)}{%
  #1\;\delimsize\|\;#2%
}

\def\tsc#1{\csdef{#1}{\textsc{\lowercase{#1}}\xspace}}
\tsc{WGM}
\tsc{QE}
\tsc{EP}
\tsc{PMS}
\tsc{BEC}
\tsc{DE}

\usepackage{tikz}
\usepackage{xstring}
\usepackage[nolist]{acronym}
\usepackage{graphicx}
\usepackage[american]{circuitikz}
\ctikzset{bipoles/resistor/width=.7}
\ctikzset{bipoles/resistor/height=.36}

\newcommand{\nomunit}[1]{%
\renewcommand{\nomentryend}{\hspace*{\fill}#1}}

\usepackage{nomencl}
\usepackage{etoolbox}
\usepackage{amssymb}
\usepackage{mathtools}
\usepackage{adjustbox}
\usepackage{placeins}
\usepackage{multirow}
\renewcommand\nomgroup[1]{%
  \item[\bfseries
  \ifstrequal{#1}{A}{Parameters}{%
  \ifstrequal{#1}{B}{Super- and subscripts}{%
  \ifstrequal{#1}{C}{Other Symbols}{}}}%
]}

\makenomenclature
\usepackage[acronym]{glossaries}

\usepackage{array}
\usepackage{makecell}

\begin{document}
\let\WriteBookmarks\relax
\def\floatpagepagefraction{1}
\def\textpagefraction{.001}
\shorttitle{Estimation of Evaporator Valve Sizes in Supermarket Refrigeration Cabinets}
\shortauthors{Leerbeck et~al.}

\title [mode = title]{Estimation of Evaporator Valve Sizes in Supermarket Refrigeration Cabinets}                      

\tnotetext[1]{This document is the results of the research
   project Digital Twins (\url{https://digitaltwins4hprs.dk/}) partly funded by the EUDP programme.}


\author[1]{K. Leerbeck}[type=editor,
                        orcid=0000-0002-1048-9503]
\cormark[1]
\ead{kenle@dtu.dk}

\credit{Writing, Original draft preparation, Data curation, Conceptualization, Methodology, Software}

\address[1]{DTU Compute, Bygning 324, 2800 Kongens Lyngby, Denmark}

\author[1]{P. Bacher}[orcid=0000-0001-5456-2576]
\ead{pbac@dtu.dk}

\credit{Conceptualization, Methodology, Supervision, Review}

\author[2]{C. Heerup}
\ead{chp@teknologisk.dk}

\credit{Conceptualization, Methodology, Supervision, Review}

\address[2]{Gregersensvej 8, 2630 Taastrup, Denmark}

\author[1]{H. Madsen}[orcid=0000-0003-0690-3713]
\ead{hmad@dtu.dk}

\credit{Review}

\cortext[cor1]{Corresponding author}
\cortext[cor2]{Principal corresponding author}

\nonumnote{In this work we demonstrate bla bla
  }

\begin{abstract}
In many applications, e.g. fault diagnostics and optimized control of supermarket refrigeration systems, it is important to determine the heat demand of the cabinets. This can easily be achieved by measuring the mass flow through each cabinet, however that is expensive and not feasible in large-scale deployments. Therefore it is important to be able to estimate the valve sizes from the monitoring data, which is typically measured. The valve size is measured by an area, which can be used to calculate mass flow through the valve -- this estimated value is referred to as the valve constant. A novel method for estimating the cabinet evaporator valve constants is proposed in the present paper. It is demonstrated using monitoring data from a refrigeration system in a supermarket in Otterup (Denmark), consisting of data sampled at a one-minute sampling rate, however it is shown that a sampling time of around 10-20 minutes is adequate for the method. Through thermodynamic analysis of a two stage CO$_2$ refrigeration system, a linear regression model for estimating valve constants was developed using time series data. The linear regression requires that transient dynamics are not present in the data, which depends on multiple factors e.g. the sampling time. If dynamics are not modelled it can be detected by a significant auto-correlation of the residuals. In order to include the dynamics in the model, an Auto-Regressive Moving Average model with eXogenous variables (ARMAX) was applied, and it is shown how it effectively eliminates the auto-correlation and provides more unbiased estimates, as well as improved the accuracy estimates. Furthermore, it is shown that the sample time has a huge impact on the valve estimates. Thus, a method for selecting the optimal sampling time is introduced. It works individually for each of the evaporators, by exploring their respective frequency spectrum. That way, a reliable estimate of the actual valve constant can be achieved for each evaporator in the system and it is shown that sampling times above 10 minutes was optimal for the analysed systems.

\end{abstract}

\begin{graphicalabstract}
\includegraphics[scale=.7]{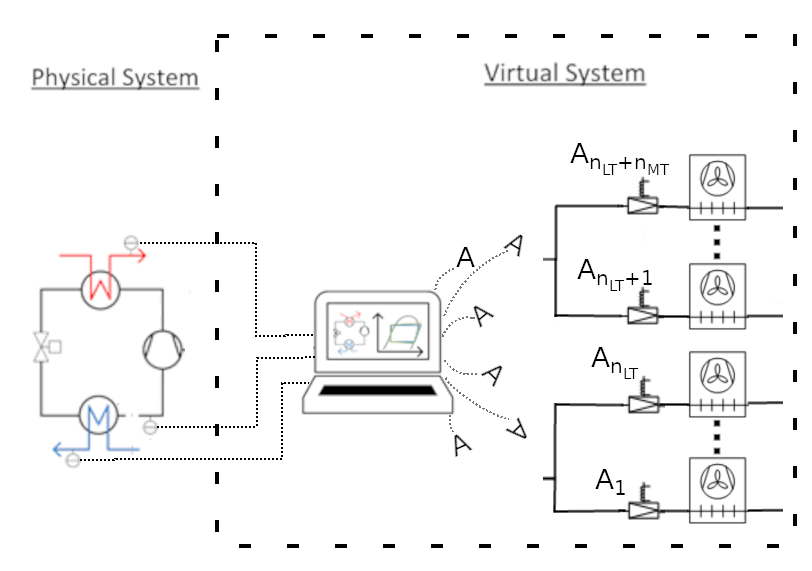}
\end{graphicalabstract}

\begin{highlights}
\item Analysis and modelling of supermarket refrigeration system.
\item Estimation of valve sizes for evaporators in refrigeration cabinets.
\item Comparison of estimation accuracy of linear regression and ARMAX models.
\end{highlights}


\maketitle
\printnomenclature

\section*{Abbreviations}

\noindent 
\begin{tabular}{@{}ll}
OLS & Ordinary Least Square\\
ARMAX & \makecell[l]{Auto-Regressive Moving Average model with \\ eXogenous variables} \\
MIMO & Multiple Inlet Multiple Outlet \\
CI & Confidence Interval

\end{tabular}

\input{Chapters/Introduction}

\input{Chapters/Methodology}

\input{Chapters/Results}
\input{Chapters/Discussion}

\input{Chapters/Conclusion}

\section{Acknowledgement \& Funding}
This document is the results of the research projects \textit{Digital twins for large-scale heat pumps and refrigeration systems} (EUDP 64019-0570) and \textit{Flexibile Energy Denmark (FED)} (IFD 8090-00069B).

\printcredits


\bibliographystyle{unsrt}
\bibliography{cas-refs}


\bio{}
\subsection*{Author biography}
	\textbf{Short-term forecasting of CO$_2$ emission intensity in power grids by machine learning}, by \textit{K. Leerbeck, P. Bacher, R. Junker, G. Goranovic, O. Corradi, R. Ebrahimy, A. Tveit, H. Madsen}, 
		published in \textit{Applied Energy} in \textit{2020}, DOI: \url{https://doi.org/10.1016/j.apenergy.2020.115527}.
		
	\textbf{Control of Heat Pumps with CO$_2$ Emission Intensity Forecasts}, by \textit{K. Leerbeck, P. Bacher, R. Junker, O. Corradi, R. Ebrahimy, A. Tveit, H. Madsen}, 
		published in \textit{Energies} in \textit{2020}, DOI: \url{https://doi.org/10.3390/en13112851}.
		
	\textbf{Grey Box Modeling of Supermarket Refrigeration Room}, by \textit{K. Leerbeck, P. Bacher, C. Heerup}, 
		published in \textit{Proceedings of the 2021 International Conference on Electrical, Computer and Energy Technologies}.

\endbio

\end{document}

%% file: Chapters/Introduction.tex
\section{Introduction}
To take advantage of system modeling and automation, the use of digital twins (a digital representation of a physical system) are increasingly evolving \citep{Yuchen,BACHER20111511}. With the increasing amount and resolution of data being collected from all kinds of sources, the potential and applications are expanded, opening up new research questions to be answered. This paper focus specifically on estimation of evaporator valve constants (a measure of the valve area when fully opened) of the evaporators in supermarket refrigeration cabinets, to enable a better understanding of these systems on a large scale. Without correct valve constants, the energy demand for each evaporator cannot be determined without measuring the mass flow directly which is expensive and practically infeasible.

Grey box modeling -- characterized by being a combination of white-box (physically based models) and black-box (purely data-driven models) -- of supermarket refrigeration systems is a research area of great interest and potential. The models can be used in for applications in optimized control and fault detection.  Thus, we achieve physical insight of the parameters of the overall and complex system using measured time series data, \citep{MARUTA20131090,KRISTENSEN20041431}. The method has been well tested in many applications, see e.g. \citep{BROK2019494,KIM2018359} for  refrigeration systems applications.
A promising approach to data-driven modeling and optimization of refrigeration systems is with neural networks and predictive control presented in \citep{KIZILKAN201111686} to increase the efficiency of the compressor in a refrigeration system. This is later followed-up in \citep{pr8091106}, where an energy reduction of 17\% was achieved on a one-stage system. However, supermarkets' refrigeration systems are usually two-stage systems with two temperature levels, one for frozen goods and one for refrigerated goods, each with its own compressor equipment, which significantly complicates the models.

Data-driven modelling of supermarket refrigeration systems can be done by modelling single components; compressor, condenser or evaporator separately \citep{Perez,Cecchinato,Willatzen} -- or the system may be modelled as a near-complete refrigeration system with several components integrated \citep{Aprea,Wang,Zhou}. Detailed data-driven models of a complete system was developed in \citep{Glenn}, using sub-space models -- a method for model parameterization of non-linear Multiple Inlet Multiple Outlet (MIMO) systems \citep{Katayama}. In that work, the cooling cabinets were not considered individually. For single component models, the parameterization can be done in a higher detail and will have fewer parameters per model compared to multi-component model.
A clever way to use both principles, is to first parameterize separate models of the main components (e.g. cooling cabinets, compressors and the condenser) -- this is done in \citep{Ehsan}, using Predictive Error Minimization for parameter estimation. Afterwards the complete system was modelled using the estimated parameters. The model was used to develop control strategies for providing energy flexibility aiming to integrate supermarket refrigeration systems in smart grids. 
  \begin{figure}[t]
	\centering
		\includegraphics[scale=.2]{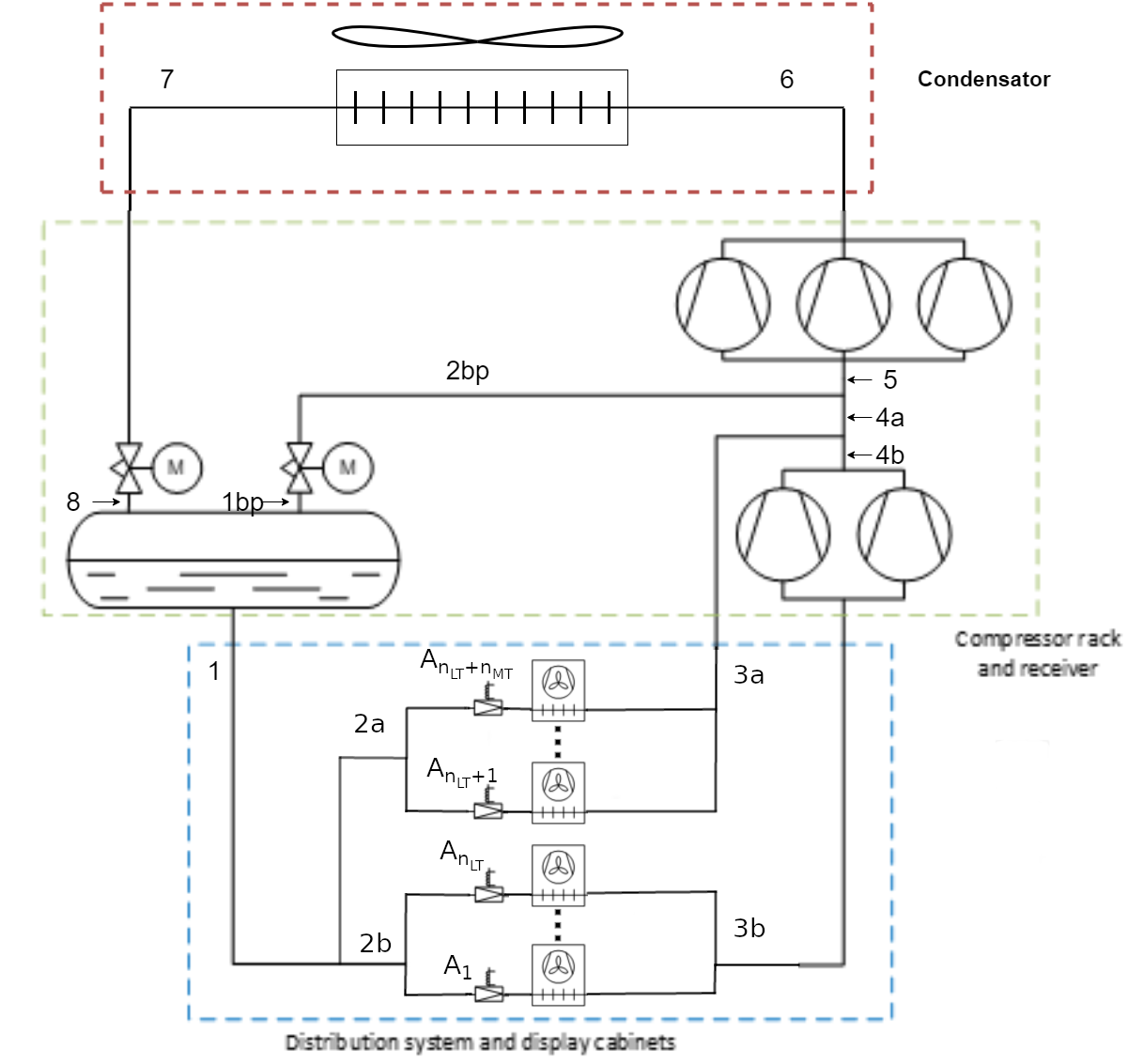}
	\caption{Exemplifying flow sheet of supermarket CO\textsubscript{2} refrigeration system. The valve constant for evaporator $i$ is denoted with $A_i$. The $n_{\text{MT}}$ and $n_{\text{LT}}$ are the number of Medium -and Low Temperature evaporators respectively. }
	\label{fig:System}
\end{figure}
Even though the models after parameter estimation may generate good predictions, there is no guarantee that the parameters are reasonable from a physical perspective. It is quite often the case that there are too many parameters to estimate, such that the models become over-parameterized and unidentifiable. In the modelling of cooling cabinets in \citep{Ehsan}, this is the case. The model can predict the temperature which is the only thing needed for control. However, if the parameters are not physically correct, we have no knowledge about the energy demand. 

The problem with many previously proposed models is their lack of practicality due to high complexity. In \cite{GreyBox}, this issue is addressed, arguing that simpler models are necessary for automatic large-scale deployment. Furthermore, the issue related to constantly changing parameters in the system model due to varying amounts of goods on store was addressed. Here, a fixed valve constant is used (known from specifications), the valve constants will in most cases be unknown for automatic large-scale deployment.
Therefore, the present paper focuses on estimating these valve constants using a data driven approach. 
First, a mass balance is formulated as a linear regression model -- with the valve constants being the model coefficients. Thus, the Ordinary Least Squares solution (OLS) is used to estimate the valve constants. Furthermore, a method for selecting the optimal sampling time is suggested and applied for each evaporator, it is based on their respective frequency spectrum.
One assumption which must hold for the linear regression model to provide unbiased estimates is that there are no transient dynamics in the modeled data. By analysing residuals it is found that they have a significant and high auto-correlation -- which means that the dynamics were not modeled. To account for this, an ARMAX (Auto Regressive Moving Average with eXogenous variables) model is applied for more accurate estimates of the valve constants.
The methods presented in this paper opens up for a fully automatized method for modelling supermarket refrigeration systems on a large scale with limited access to knowledge of the individual supermarkets, only monitoring data, which is typically available, is used.

%% file: Chapters/Methodology.tex
\nomenclature[A]{$\dot{m}$}{Mass flow 
  \nomunit{$\text{kg}/\text{s}$}}
\nomenclature[A]{$P$}{Pressure
  \nomunit{Bar}}
\nomenclature[B]{rec}{Receiver
  \nomunit{}}
\nomenclature[B]{$e$}{Evaporator
  \nomunit{}}
\nomenclature[B]{liq}{Liquid phase
  \nomunit{}}
\nomenclature[A]{$\rho$}{Density
  \nomunit{kg/m$^3$}} 
 \nomenclature[A]{$\lambda$}{Opening degree (0-1)
  \nomunit{-}}
 \nomenclature[A]{$A$}{Valve constant
  \nomunit{$\sqrt{10}$ m$^2$}}
 \nomenclature[A]{$A$}{Valve constant
  \nomunit{$\sqrt{10}$ m$^2$}}
 \nomenclature[A]{f}{Virtual compressor frequency (load as percentage of maximum capacity 0-1)
  \nomunit{$-$}}
\nomenclature[A]{$\eta_{\text{vol}}$}{Volumetric efficiency (0-1)
  \nomunit{$-$}}
 \nomenclature[A]{$V_\text{s}$}{Compressor stroke volume
  \nomunit{m$^3$}}
 \nomenclature[B]{gas}{Gas phase
  \nomunit{}}  
 \nomenclature[A]{$Q$}{Gas quality (0-1)
  \nomunit{-}}  
  \nomenclature[A]{$h$}{Enthalpy
  \nomunit{J/kg}} 
 \nomenclature[B]{LT}{Low temperature
  \nomunit{}} 
 \nomenclature[B]{MT}{Medium temperature
  \nomunit{}}
  \nomenclature[B]{comp}{Compressor rack
  \nomunit{}}
  \nomenclature[B]{gc}{Gas cooler
  \nomunit{}}

\section{The system}
The refrigeration system under study is illustrated in Figure \ref{fig:System}. Temperature and pressure measurements are accessible at all numbered points on the figure. The process is the following. After the receiver, at stage "1", the refrigerant is liquefied and split into the medium temperature (MT) string and the low temperature (LT) string, where expansion valves drops the pressure to the desired saturation temperature. The evaporators are controlled with valves using either hysteresis (open/close) or modulating (continues open) control -- valve $i$ has a corresponding valve constant, denoted $A_i$,. After the evaporators at stage "3a" and "3b", the refrigerant is superheated to avoid any droplets from entering the compressor. At stage "4a", "4b" and "5", after the low pressure compressor rack, the pressure is the same, but the enthalpy varies as the MT string and by-pass (bp) string from the receiver connects. The refrigerant now enters the high pressure compressor rack and continues through the condenser to a sub-cooled state throttled to stage ”8” and split in a liquid and a gas fraction  -- and the cycle repeats. Note, at higher ambient temperatures the refrigerant does not condense, but the gas is throttled and this result in a lower liquid and a higher gas fraction at “8”.

This research focus on modelling the evaporators, hence anything between stage "6" and "8" will not be discussed further.

\section{Methodology}

In this section, first, under the assumption that there are no transient dynamics in processes, the steps in setting up a linear regression model for estimation of the valve constant are described, and second, the applied OLS estimation technique is presented, and finally to include linear dynamics an ARMAX model \cite{madsen2007time} is presented.

The valve constant $A_i$ is used to determine the refrigerant mass flow through evaporator $i$ by
\begin{align}\label{eq:mass_cab}
    \dot{m}_{\text{cab},i,t} = \sqrt{(P_{\text{rec},t} - P_{\text{e},i,t} ) \rho_{\text{liq},t}  } \lambda_{i,t} A_i,
\end{align}
where $P_{\text{rec},t}$ is the receiver pressure at time $t$ and $\rho_{\text{liq},t}$ is the density of the liquefied refrigerant, which is found as a function of $P_{\text{rec},t}$ using CoolProp \cite{coolprop}. For the $i$'th cabinet $P_{\text{e},i,t}$ is the evaporator pressure at time $t$, $\lambda_{i,t}$ is the opening degree of the valve at time $t$, given as a value between 0 and 1 and is controlled either using hysteresis (open/close) or modular (continues open) control. Hence, the valve constant, $A_i$, is defined as an area equivalent to the area of the valve opening when the valve is fully open.

The mass balance at stage "4a" (Figure \ref{fig:System}) can be written as
\begin{align}\label{eq:mass_balance}
    \dot{m}_{\text{MT,comp,t}} - \dot{m}_{\text{bp},t} = \sum_{i=1}^{n_{\text{LT}}} \dot{m}_{\text{cab},i,t} + \sum_{i=n_{\text{LT}}+1}^{n_{\text{MT}}} \dot{m}_{\text{cab},i,t},
\end{align}
where the $\dot{m}_{\text{MT,comp},t}$ is the mass flow going through the medium pressure (temperature) compressor at time $t$ and $\dot{m}_{\text{bp},t}$ is the by-pass mass flow entering from the receiver at time $t$. The right side mass flows are simply the sum of all the mass flows through the LT and MT cabinets, where $n_{\text{LT}}$ and $n_{\text{MT}}$ are the total number of LT and MT cabinets, respectively.

The mass flows on the left side of Equation \ref{eq:mass_balance} are calculated by
\begin{align}\label{eq:mass_MT}
    \dot{m}_{\text{MT,comp},t} = f_{\text{comp},t} \eta_{\text{vol}}  V_\text{s} \rho_{\text{gas}},
\end{align}
where f$_\text{comp}$ is the virtual compressor frequency (load as percentage of maximum capacity) at time $t$, $\eta_{\text{vol}} \sim 0.9 $, $V_\text{s}$ is the displacement volume which is specified for the compressor, and finally the mass flow from the by-pass valve is calculated by
\begin{align}\label{eq:mass_bp}
    \dot{m}_{\text{bp},t} = Q_t (h_{\text{gc},t}, P_{\text{rec},t})\, \dot{m}_{\text{MT,comp},t}.
\end{align}
where $Q_t$ is the gas quality of the refrigerant from the by-pass valve as a function of the enthalpy in the gas cooler, $h_{\text{gc},t}$, and the receiver pressure $P_{\text{rec},t}$ at time $t$.

In order to set up the a linear regression model with the valve constants as coefficients, we let
\begin{align}
x_{i,t} =  \sqrt{(P_{\text{rec},t} - P_{\text{e},i,t}) \rho_{\text{liquid},t}} \lambda_{i,t},
\end{align}
such that the right side of Equation \ref{eq:mass_balance} can be rewritten as
\begin{align}
  \sum_{i=1}^{n_{\text{LT}}} x_{i,t} A_i + \sum_{i=n_{\text{LT}}+1}^{n_{\text{MT}}} x_{i,t} A_i,
\end{align}
and
\begin{align}
    y_t = \dot{m}_{\text{MT,comp},t} - \dot{m}_{\text{bp},t}.
\end{align}
The linear regression model can by set up by
\begin{align}
    \boldsymbol{y} &= \begin{bmatrix}
                        y_{1} & y_{2} & \cdots &  y_{n}
                    \end{bmatrix}^T \nonumber \\
    \boldsymbol{x} &= \begin{bmatrix}
                        x_{1,1} & x_{2,1} &  \cdots & x_{n_\text{V},1} \\
                        x_{1,2} & x_{2,2} & \cdots & x_{n_\text{V},2}\\
                        \vdots & \vdots &   \ddots & \vdots \\
                        x_{1,n} & x_{2,n} & \cdots & x_{n_\text{V},n} 
                    \end{bmatrix}\\
    \boldsymbol{\beta} &= \begin{bmatrix}
                        A_{1} & A_{2} &  \cdots & A_{n_V} 
                    \end{bmatrix}^T ,
                    \nonumber 
\end{align}
where $n$ is the number of observations and $n_\text{V} = n_{\text{MT}} + n_{\text{LT}}$, and thus can be written as
\begin{align}
    \boldsymbol{Y} = \boldsymbol{x}\boldsymbol{\beta} + \boldsymbol{\epsilon},
\end{align}
where the errors are random variables in the vector  
\begin{align}
  \boldsymbol{\epsilon} = \begin{bmatrix}
                        \epsilon_{1} & \epsilon_{2} & \cdots &  \epsilon_{n},
                    \end{bmatrix}^T
\end{align}
which must fulfill $\epsilon_t \sim N(0,\sigma^2 I)$ and i.i.d. Note, that $\boldsymbol{Y}$ is capital, since it is the model output and thus a vector of random variables.

The parameter vector $\boldsymbol{\beta}$ can be estimated from the OLS regression that minimize
\begin{align}
S(\boldsymbol{\beta}) =  ||\boldsymbol{Y} - \boldsymbol{x}\boldsymbol{\beta}||^2 
\end{align}
which has the OLS solution
\begin{align}
\hat{\boldsymbol{\beta}} = (\boldsymbol{x}^T \boldsymbol{x})^{-1} \boldsymbol{x}^T \boldsymbol{y}.\label{eq:ols}
\end{align}

Furthermore, modelling time series variables with a linear regression model will, if there are any dynamics present, lead to auto-correlation of the errors. This violates the assumption of i.i.d and consequently, the coefficient estimates can be biased and statistics not valid. Thus, to account for this, we will consider an ARMAX model. Using the back-shift operator, $B$, this model can be written as
\begin{equation}
    Y_t = \boldsymbol{x}_t\boldsymbol{\beta} +  \frac{\theta(B)}{ \phi(B)} \epsilon_t
\end{equation}
where $\theta(B) = 1 + \theta_1 B^1 + ... + \theta_q B^q $ and $\phi(B) = 1 + \phi_1 B^1 + ... + \phi_p B^p$. 
 The model is refered to as an ARMAX\textsubscript{p,q} model, where parameters $p$ and $q$ determines the order of the AR and MA part, respectively. Note, that there are no lags in input variable, $x_t$, in this formulation, hence it is not a full ARMAX model since it is constrained to a model order of the input polynomial set to zero, i.e.\ $s=0$ according to \cite{TSA}. For the parameter estimation, the \textsf{R} function \texttt{arimax} from the \texttt{TSA} package is used. It is estimating the parameters using based the maximum likelihood method.

\begin{figure*}[t]
  \centering
  \begin{minipage}[b]{0.45\textwidth}
    \includegraphics[width=\textwidth]{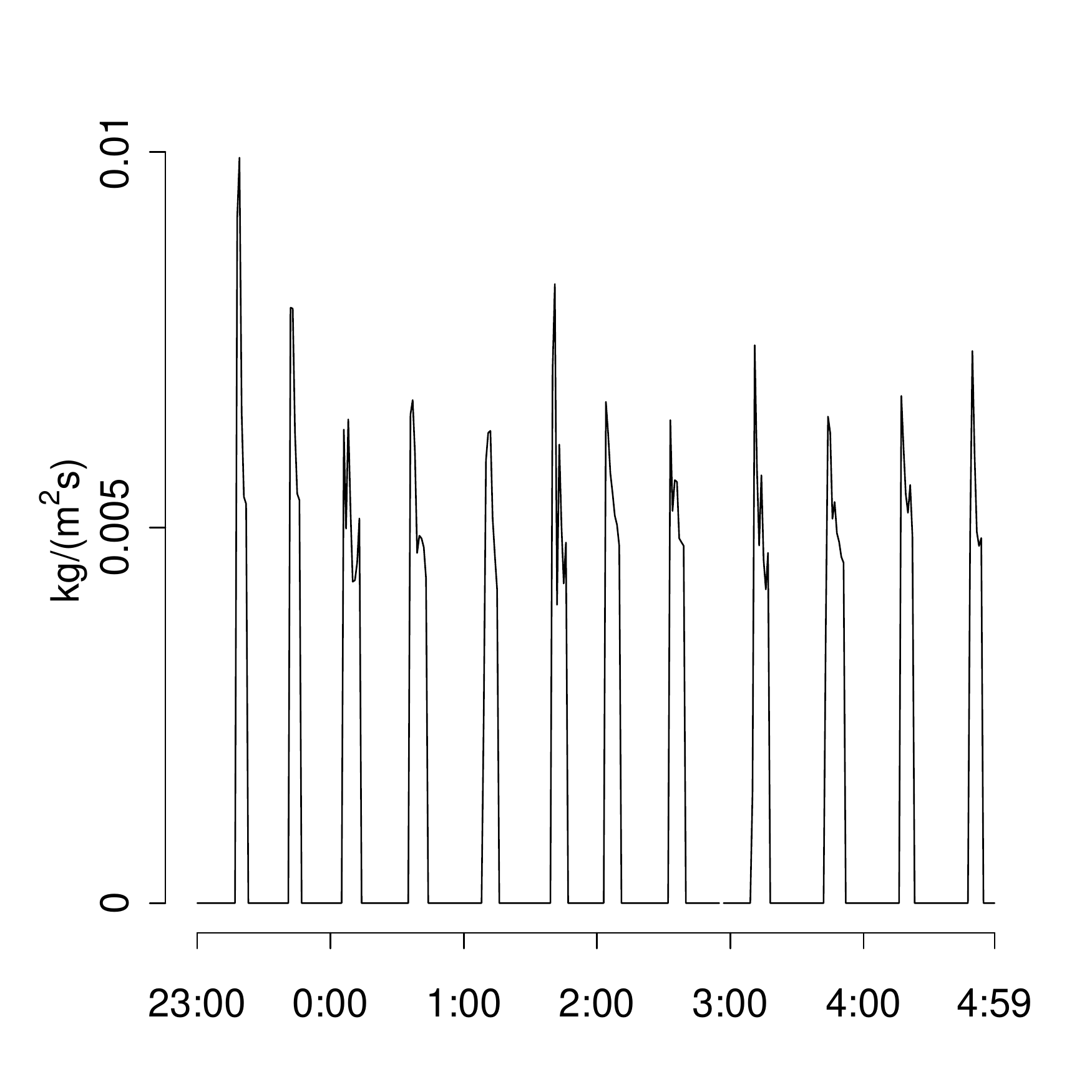}
    \caption{$x_{i,t}$ of the cooling room MT$_2$ the 29th of October 2013.}\label{fig:night}
  \end{minipage}
  \hfill
  \begin{minipage}[b]{0.45\textwidth}
    \includegraphics[width=\textwidth]{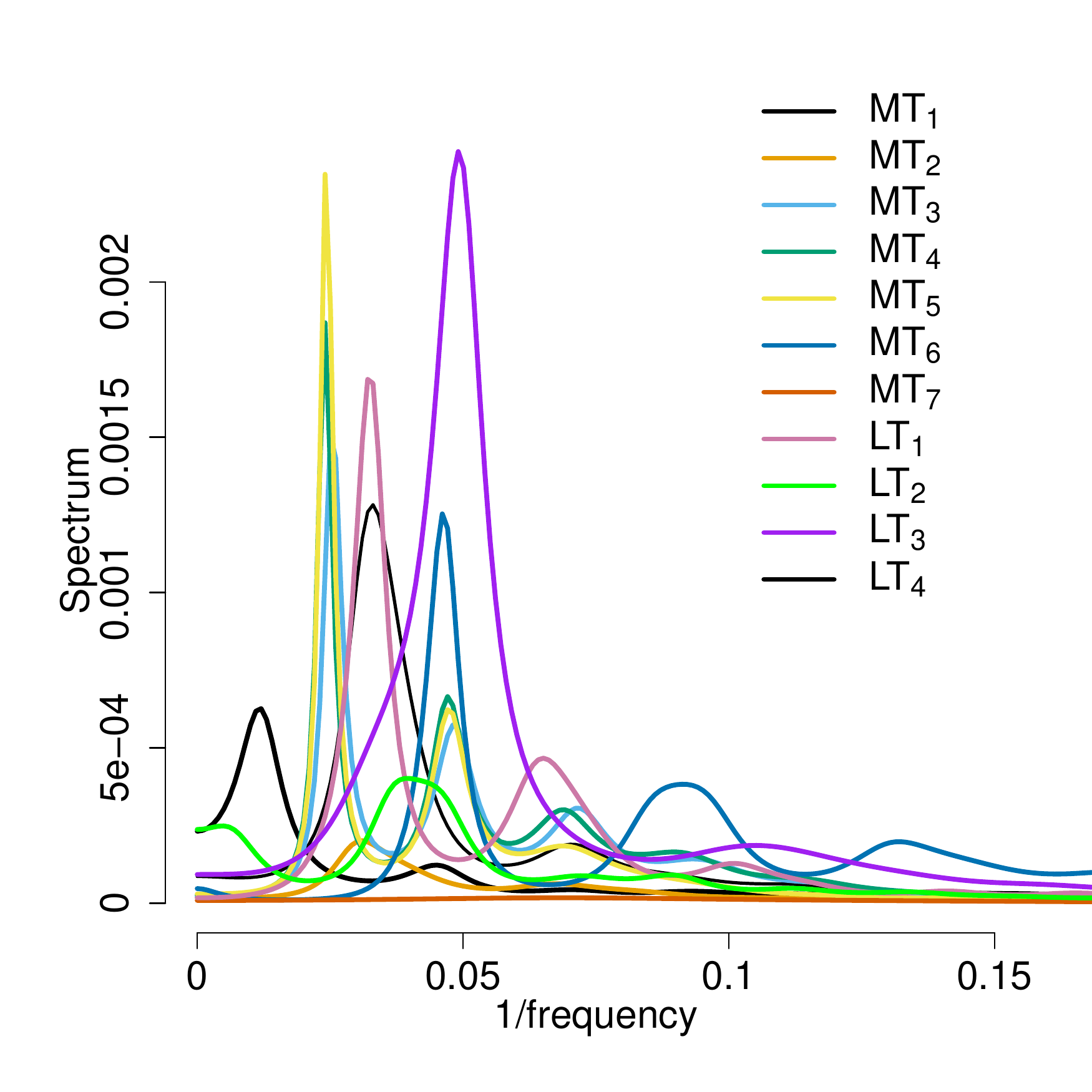}
    \caption{The frequency spectrum for each of the evaporators.}\label{fig:specs}
  \end{minipage}
\end{figure*}

\section{Case Study}\label{sec:data}
\subsection{Data and sampling time}
The data used consists of 1-minute time series covering an eight months period from November 2013 to June 2014. The values for one cabinet during one night is illustrated in Figure \ref{fig:night}. If the raw 1-minute data is used, a lot of redundant high frequency dynamics are present in the data that will lead to a poor model with high auto correlation. Thus we need to explore the impact of the sampling time on the model.
From the frequency spectrum shown in Figure \ref{fig:specs} we observe that most of the cabinets have a clear peak. We can extract the inverse frequency at the peak for each cabinet -- that is the most often occurring cycle time for the valve opening. The peak cycle times identified are listed in Table \ref{tab:frequencies}. Note, cabinet MT$_7$ has an extremely flat curve compared to all the other evaporators, hence its 14 min. cycle time is not as significant. This is because it uses modulating control (continuously valve opening) thus it has no clear peak in the frequency spectrum. Furthermore, it is noticed that MT$_1$ has the largest cycle time of 83 minutes. However, its spectrum peak is low compared to the other evaporators -- it is a booster evaporator for MT$_2$ (they are both located in the refrigeration room) and continuously supplies the cooling that MT$_2$ cannot. The rest of the evaporators have cycles varying from 20 min. to 40 min., where MT$_3$, MT$_4$ and MT$_5$ are clustered around 40 min. -- these are three different evaporators in the same cooling cabinet. For a good model, we will need to have at least, and ideally, two measurements for each cycle (one open and one closed). Because the evaporators have such different cycle times, there is no sampling time that will fit all the evaporators optimally. For instance, a 20-minute sampling time is best for MT$_3$, MT$_4$ and MT$_5$, whereas a 10-minute sampling time is best for MT$_6$, LT$_2$ and LT$_3$. For this study, we will therefore examine sampling time in the range of 1 min. to 30 min. The re-sampling from 1-minute sampling time to higher sampling times is carried out by averaging in the time periods, such that aliasing is avoided and a better signal-to-noise ratio in the re-sampled data is obtained \cite{TSA}. 

In order to evaluate if the estimated valve constants, $A_i$, have realistic values they are compared to the specified valve constants \cite{Fredslund} (see Table \ref{tab:danfoss_valves}).

\begin{table}[htbp]
\small
  \centering
    \begin{tabular}{p{1cm} p{1cm} p{1cm} p{1cm} p{1cm} p{1cm} }
    MT$_1$  & MT$_2$  & MT$_3$  & MT$_4$  & MT$_5$   & MT$_6$   \\
    \hline 
    83.17 & 32.19 & 39.92 & 41.58 & 41.58 & 21.23 \\[10pt]
     MT$_7$  & LT$_1$  & LT$_2$  & LT$_3$  & LT$_4$  & \\
     \hline
     14.68 & 31.19 & 25.59 & 20.37 & 30.24 &

\end{tabular}%
\caption{The dominant inverse frequency for each cabinet in minutes.}\label{tab:frequencies}
\end{table}%

\begin{figure*}[t]
	\centering
		\includegraphics[scale=.35]{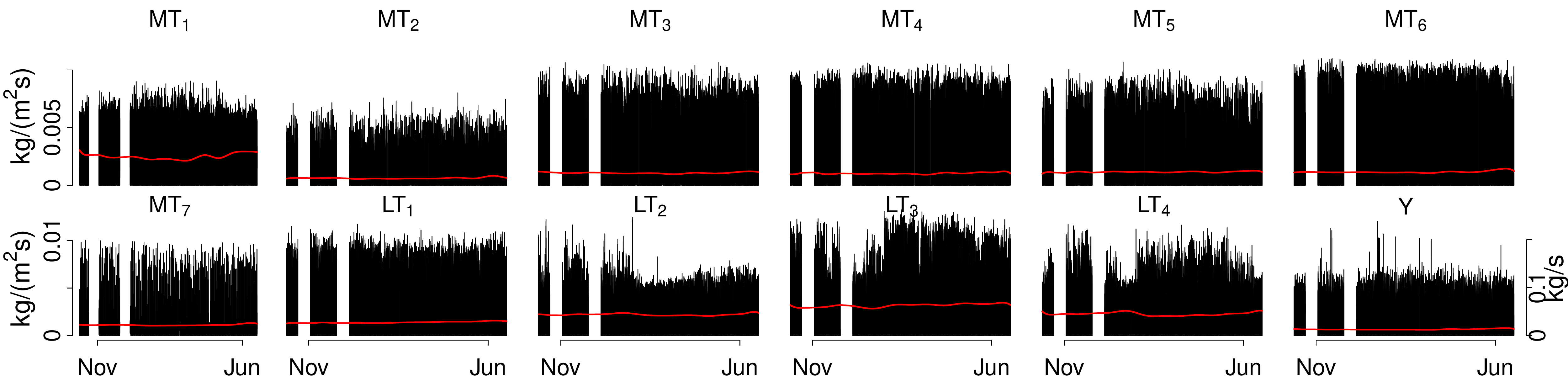}
	\caption{Daily averages of $x_{i,t}$ for each cabinet and $\mathbf{Y}$.}
	\label{fig:timeseries}
\end{figure*}
\begin{figure*}[t]
  \centering
    \includegraphics[scale=.35]{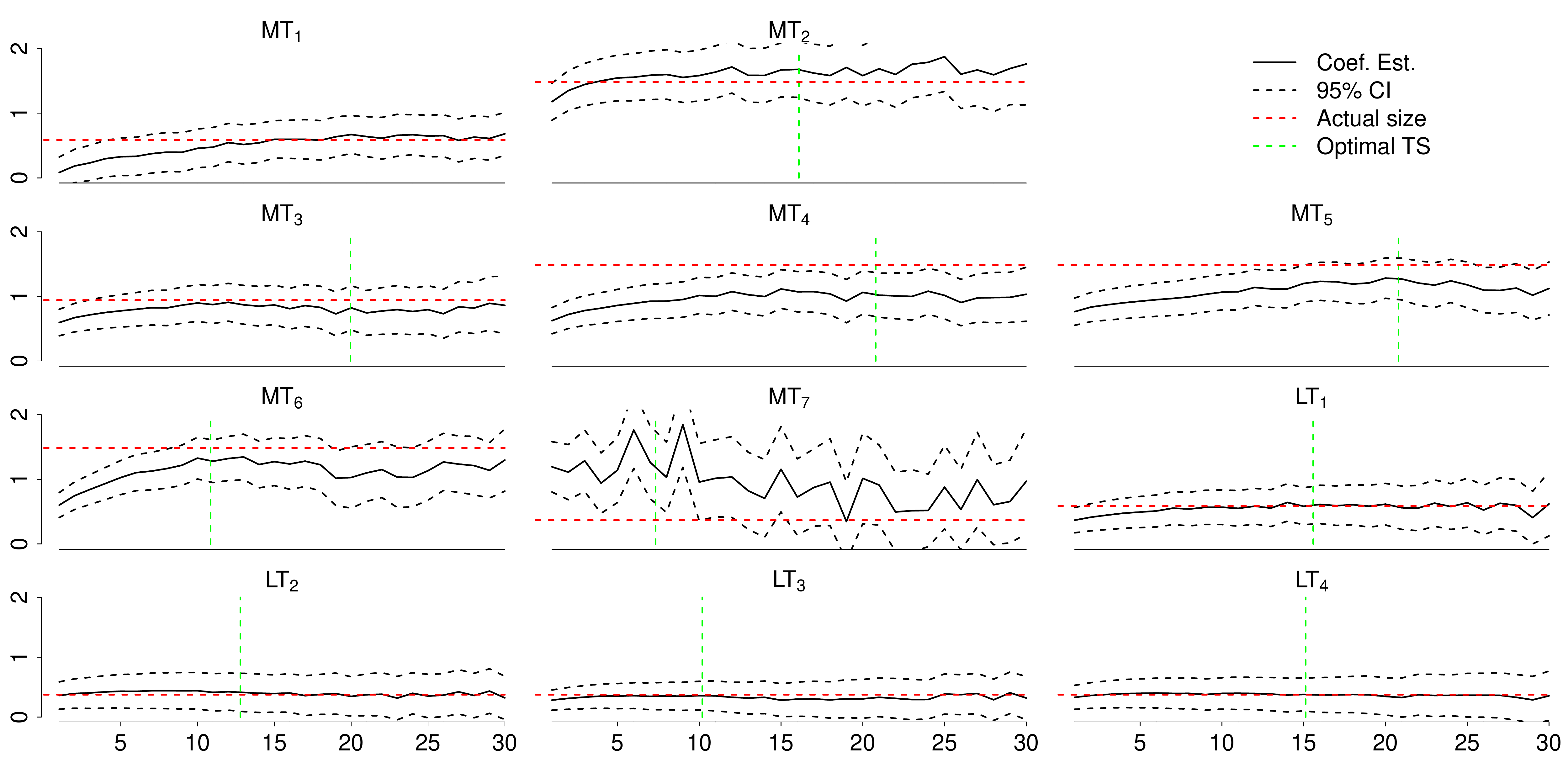}
	\caption{Estimated valve constant, $A_i$, for evaporator MT$_i$ as a function of the sampling time using all the data -- November 2013 to June 2014.}\label{fig:res}
\end{figure*}


The operation of evaporators within the same cabinet is obviously quite similar, thus correlations might occur -- this can be crucial for the model, as this can lead to problems with co-linearity in the estimation. When this happens the condition number of the matrix $\mathbf{X}$ is lowered, making the determinant of it close to zero, and thus inverting it results in large numerical errors, see Equation \ref{eq:ols}. However, in the present case the co-linearity was not an issue (generally, a correlation coefficient $>0.7$ indicates co-linearity \citep{Dormann} ). To confirm further, the Variance Inflation Factor (VIF) is a widely used tool -- a variant of stepwise regression used to detect multi co-linearity. A VIF less than 10 is an acceptable level \citep{Hair}, in present case the VIF was less than 2.


%% file: Chapters/Results.tex
\section{Results}

As discussed in Section \ref{sec:data}, the sampling time potentially impacts the estimated parameters, thus this was investigated and the result of this is presented in this section -- and how the suggested method for determining the optimal time resolution, identified based on their cycle frequencies, see Figure \ref{fig:specs} and Table \ref{tab:frequencies}, lead to reliable estimates. First the results of applying the linear regression model are presented and thereafter the results from the ARMAX model.

\subsection{Linear Regression Model}

\begin{figure*}
 \includegraphics[scale=0.75]{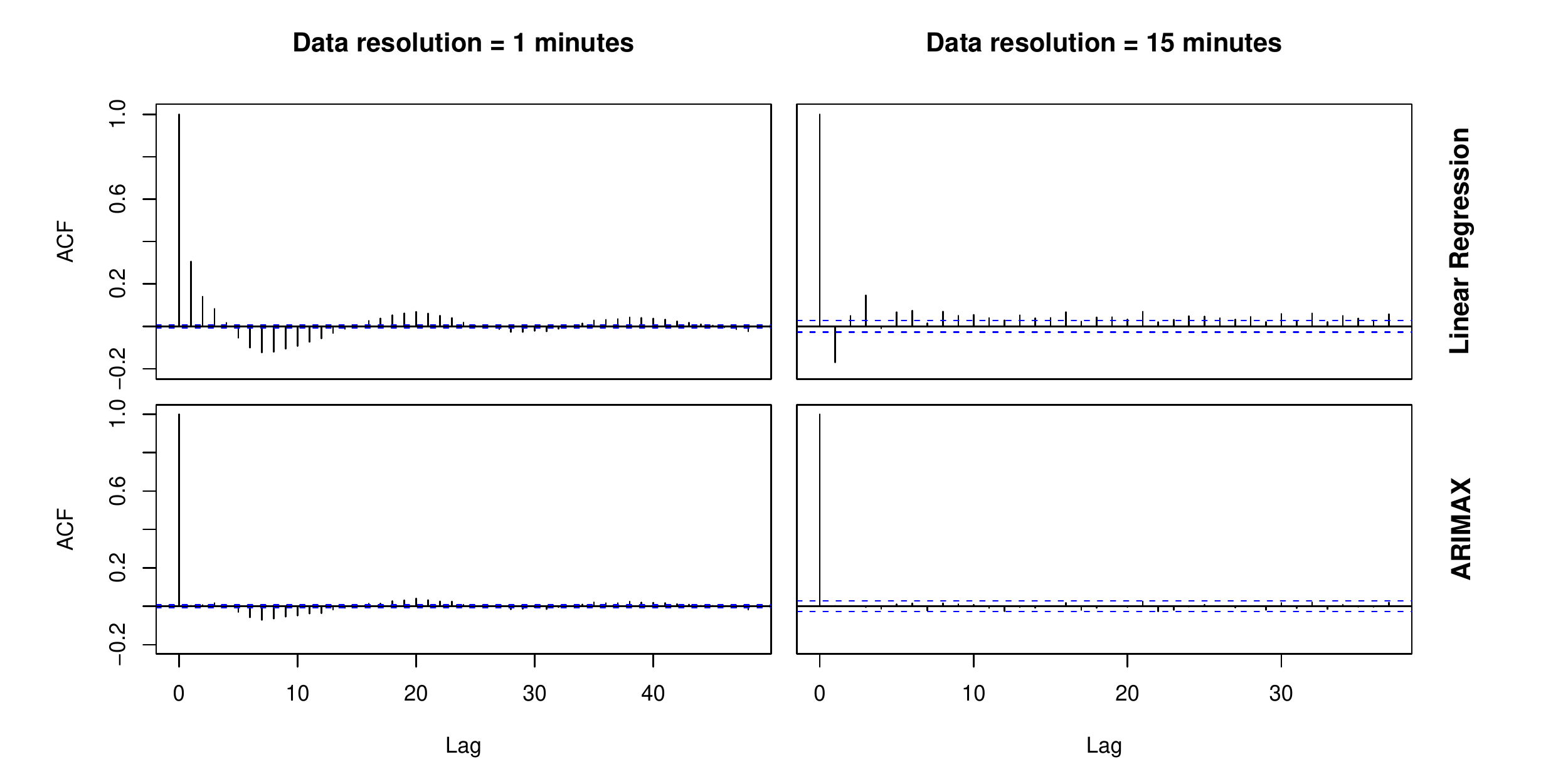}
	\caption{Auto Correlation functions (ACF) for the linear regression models and ARMAX models using 1 minute and 15 minute sampling time.}\label{fig:acf}
\end{figure*}

The estimated valve constants are presented as function of the sampling time in Figure \ref{fig:res}. In each figure the red line marks the specified valve constant (see Table \ref{tab:danfoss_valves}), the green line marks the identified optimal sampling time and the black line is the valve constant estimates (with the 95\% CI) vs. the sampling time from 1 min. to 30 min.. The estimates clearly vary with the sampling time, however the estimator converge to a constant level with higher sampling times in most cases. Both the estimates for MT$_2$ and MT$_5$ seem to decrease after the sampling time exceeds their respective optimal sampling time. Furthermore, at low sampling times (< 10 minutes) the estimates are typically diverging. By observing the Auto Correlation Function (ACF) of the residuals for the 1-minute model and the 15-minute model presented in Figure \ref{fig:acf} (top left and top right plot respectively), there is clearly significant auto-correlation for both models, which indicates the need for modelling dynamics with an ARMAX model.

\begin{table}[htbp]
\small
  \centering
    \begin{tabular}{l c c}
    Cabinet & Valve  &  $A [\sqrt{10} m^2]$ \\

    MT$_1$ & AKV 10-3  & 0.58765 \\
    MT$_2$ & AKV 10-5 &  1.48523\\
    MT$_3$ & AKV 10-4 &  0.94185 \\
    MT$_4$ & AKV 10-5 &  1.48523\\
    MT$_5$ & AKV 10-5 &  1.48523\\
    MT$_6$ & AKV 10-5 &  1.48523 \\
    MT$_7$ & AKV 10-2 &  0.37191\\
    LT$_1$ & AKV 10-3 &  0.58765\\
    LT$_2$ & AKV 10-2 &  0.37191\\
    LT$_3$ & AKV 10-2 &  0.37191\\
    LT$_4$ & AKV 10-2 &  0.37191

\end{tabular}%
\caption{Valves installed in Fakta Otterup and their valve constants $A$ - derived from a previous study \cite{Fredslund} }\label{tab:danfoss_valves}
\end{table}%

\subsection{ARMAX Model}

\begin{figure*}[t]
  \centering
    \includegraphics[scale=.35]{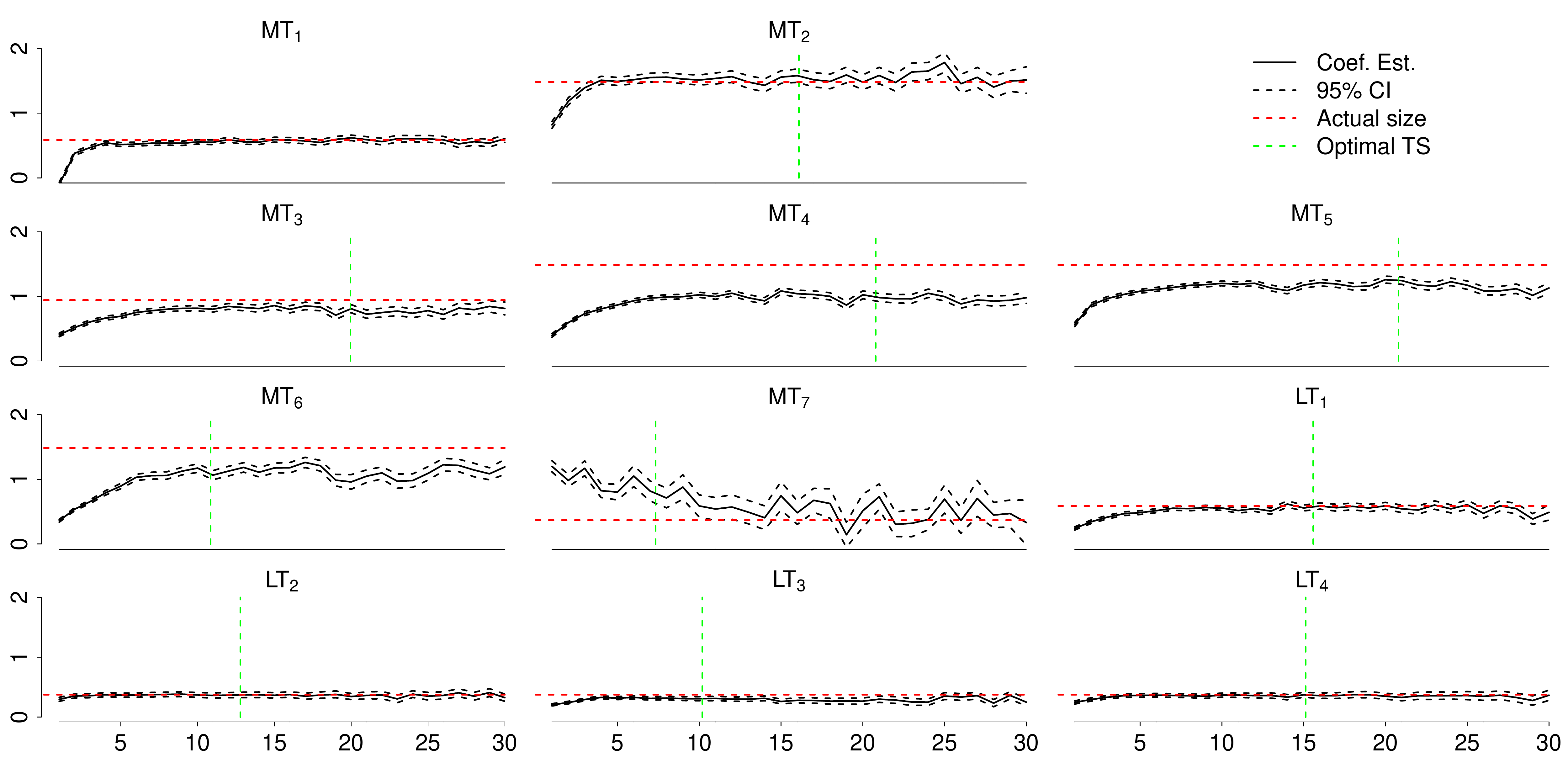}
	\caption{Estimated parameters as a function of the sampling time using all the data - November 2013 to June 2014.}\label{fig:res_arimax}
\end{figure*}

The ACF of the linear regression model using 1-minute sampling time, see Figure \ref{fig:acf} top left plot, shows seasonality caused by hysteresis control with roughly 15 lags between the wave peaks. Now, consider the ACF of the 15-minute sampling time model (Figure \ref{fig:acf}, top right), there is a large negative correlation to lag 1, which implies the need for a moving average term. The positive correlation of lag 4 implies the need for auto-regressive terms.
An ARMAX$_{4,1}$ model removed all auto-correlation in the 15 min. sampling time model (Figure \ref{fig:acf}, bottom right). In this study, an ARMAX$_{4,1}$ will be used for all sampling times, though the best ARMAX fit may be different for other sampling times.
The 1-min. model still has some seasonal auto-correlation (Figure \ref{fig:acf}, bottom left), but it is much improved. Considering the resulting valve constant estimates, presented in Figure \ref{fig:res_arimax}, firstly, it is found that the CIs are much more narrow compared with the CIs for the linear regression model. Secondly, the estimates vary with the sampling time and above a certain level the estimates converge.

%% file: Chapters/Discussion.tex
\section{Discussion}

The method for valve estimation provides useful results, however, the sampling time turned out to have a huge impact which must be handled to achieve reliable estimates. An issue with the method is that the evaporators have very different optimal sampling times given their different frequencies. Thus, there exist no single sampling time suitable for all the evaporators. However, an approach for identifying an optimal sampling time for each evaporator was presented and is was shown that estimates, at their respective optimal sampling time, fits fairly well with the specified valve constants. Notice, for the linear regression model the optimal sampling time is at a point on the graphs (in Figure \ref{fig:res}), where the valve constant estimates either begin to converge, or they are found in a local optimum. For the ARMAX model, the sampling time is far less important -- the valve constant estimates do not vary much using sampling times above $10$ min.. However, we still see generally more accurate results around the optimal sampling time. These effects should be studied in further research including multiple supermarkets.

Models using sampling times below $10$ min. provided generally very different results and the residuals were also highly correlated (See Figure \ref{fig:acf} left plots for the 1 minute models) -- both in the linear regression model as well as the ARMAX$_{4,1}$ model. For models using sampling times $> 15$ min., the results were much more uniform in both the linear regression model and the ARMAX model. Problems with the residuals are however still present for the linear regression model -- these are fixed in the ARMAX$_{4,1}$ model. 

It is clear that the CIs are much more narrow for the ARMAX$_{4,1}$ models, suggesting a generally better model. The estimates for the ARMAX$_{4,1}$ models using a sampling time above $15$ min. are also less dependent on the sampling time than for the linear regression model. However, for the cabinet MT$_7$, there are still some rather large estimate variations. This is because the cabinet uses modulating control rather than hysteresis control as all the others use. Therefore, the more accurate valve constant estimates may be an average of the estimates from the ARMAX$_{4,1}$ models using sampling times between 15 and 20 min. to even out the variation. We may calculate an average of the estimates from models using sampling times around the optimal sampling time, if this exists. 

The method has only been tested on one refrigeration system with mixed controls -- one evaporator used modulating -- and the rest used hysteresis control. It will be interesting to expand the study to include supermarkets that only use modulating control and supermarkets that only use hysteresis control. Consistency among the evaporators is expected to give better results. 

This study used 8 months of data for the results -- it is, however, possible to calculate the estimates with less data, but the uncertainty will increase. As much data as possible should be used for the estimates, however, periods where faults have occurred should be removed to ensure right estimates.

%% file: Chapters/Conclusion.tex
\section{Conclusion}

Models for estimating the valve constants in supermarket refrigeration systems are presented and results are obtained from data collected in a regular supermarket in Denmark. A mass balance is formulated leading to a linear regression model from which the valve constants can be estimated -- and it is extended to an ARMAX model, which can describe the system dynamics. The estimated results are analysed and it is shown that reliable values can be obtained, but that several factors must be taken into account. Firstly, co-linearity in the inputs can cause issues, but this is not an issue in the case study -- potentially it can be a problem for other systems, and should be checked e.g. with the Variance Inflation Factor. Furthermore, the coefficient estimates vary with the sampling time and it is shown that no single sampling time is optimal for all evaporators, because they have different cycle times. However, for models using sampling times above 15 min., the valve estimates become more stable and less dependent on the sampling time. Some variation is still found, hence some smoothing of the estimates as a function of the sampling time can be beneficial, however an approach to this must be developed with data from multiple supermarkets. The suggested approach for an optimal sampling time is shown to be a good indicator for the best estimates.
Auto-correlation is highly present in the linear regression models, thus to account for this, an ARMAX$_{4,1}$ is applied. After applying the ARMAX$_{4,1}$ model, the CI for the valve estimates becomes more narrow, suggesting a more accurate model. Furthermore, the estimates are also slightly less dependent on the sampling time than with the linear regression model. Thus it is found, that the ARMAX model is preferable over the linear regression model, however still studies including a wide range of supermarkets must be carried out to further develop the methodology and study its full potential.

%% file: cas-dc-template.bbl
\begin{thebibliography}{10}

\bibitem{Yuchen}
Jiang Yuchen, Yin Shen, Li~Kuan, Luo Hao, and Kaynak Okyay.
\newblock Industrial applications of digital twins.
\newblock {\em Phil. Trans. R. Soc.}, 379, 2021.

\bibitem{BACHER20111511}
Peder Bacher and Henrik Madsen.
\newblock Identifying suitable models for the heat dynamics of buildings.
\newblock {\em Energy and Buildings}, 43(7):1511--1522, 2011.

\bibitem{MARUTA20131090}
Ichiro Maruta and Toshiharu Sugie.
\newblock Projection-based identification algorithm for grey-box
  continuous-time models.
\newblock {\em Systems \& Control Letters}, 62(11):1090--1097, 2013.

\bibitem{KRISTENSEN20041431}
Niels~Rode Kristensen, Henrik Madsen, and Sten~Bay Jørgensen.
\newblock A method for systematic improvement of stochastic grey-box models.
\newblock {\em Computers \& Chemical Engineering}, 28(8):1431--1449, 2004.

\bibitem{BROK2019494}
Niclas~Brabrand Brok, Thomas Munk-Nielsen, Henrik Madsen, and Peter~A.
  Stentoft.
\newblock Flexible control of wastewater aeration for cost-efficient,
  sustainable treatment.
\newblock {\em IFAC-PapersOnLine}, 52(4):494--499, 2019.
\newblock IFAC Workshop on Control of Smart Grid and Renewable Energy Systems
  CSGRES 2019.

\bibitem{KIM2018359}
Donghun Kim, Jie Cai, James~E. Braun, and Kartik~B. Ariyur.
\newblock System identification for building thermal systems under the presence
  of unmeasured disturbances in closed loop operation: Theoretical analysis and
  application.
\newblock {\em Energy and Buildings}, 167:359--369, 2018.

\bibitem{KIZILKAN201111686}
Önder Kizilkan.
\newblock Thermodynamic analysis of variable speed refrigeration system using
  artificial neural networks.
\newblock {\em Expert Systems with Applications}, 38(9):11686--11692, 2011.

\bibitem{pr8091106}
Josep Cirera, Jesus~A. Carino, Daniel Zurita, and Juan~A. Ortega.
\newblock Improving the energy efficiency of industrial refrigeration systems
  by means of data-driven load management.
\newblock {\em Processes}, 8(9), 2020.

\bibitem{Perez}
C.D. Pérez-Segarra, J.~Rigola, M.~Sòria, and A.~Oliva.
\newblock Detailed thermodynamic characterization of hermetic reciprocating
  compressors.
\newblock {\em International Journal of Refrigeration}, 28:579--593, 2005.

\bibitem{Cecchinato}
M.~Corradi, L.~Cecchinato, G.~Schiochet, and C.~Zilio.
\newblock Modelling fin-and-tube gas-cooler for transcritical carbon dioxide
  cycles.
\newblock In {\em International Refrigeration and Air conditioning}, 2006.

\bibitem{Willatzen}
M.~Willatzen, N.B.O.L. Pettit, and L.~Ploug-Sørensen.
\newblock A general dynamic simulation model for evaporators and condensers in
  refrigeration. part i: moving-boundary formulation of two-phase flows with
  heat exchange.
\newblock {\em International Journal of Refrigeration}, 21:398--403, 1998.

\bibitem{Aprea}
C.~Aprea and C.~Renno.
\newblock An experimental analysis of a thermodynamic model of a vapor
  compression refrigeration plant on varying the compressor speed.
\newblock {\em International Journal of Energy Research}, 28:537--549, 2004.

\bibitem{Wang}
F.~Wang, G.~Maidment, J.~Missenden, and R.~Tozer.
\newblock The novel use of phase change materials in refrigeration plant. part
  2: Dynamic simulation model for the combined system.
\newblock {\em Applied Thermal Engineering}, 27:2902--2910, 2007.

\bibitem{Zhou}
R.~Zhou, T.~Zhang, J.~Catano, J.~T. Wen, G.~J Michna, Y.~Peles, and M.~K.
  Jensen.
\newblock The steady-state modeling and optimization of a refrigeration system
  for high heat flux removal.
\newblock {\em Applied Thermal Engineering}, 30:2347--2356, 2010.

\bibitem{Glenn}
G.~Andreasen, J.~Stoustrup, R.~Izadi-Zamanabadi, and A.~Pardiñas, Á.;~Hafner.
\newblock Data-driven modeling of a co2 refrigeration system.
\newblock In {\em American Control Conference (ACC)}, 2019.

\bibitem{Katayama}
T.~Katayama.
\newblock {\em Subspace Methods for System Identification}.
\newblock Springer, 2005.

\bibitem{Ehsan}
E.~Shafiei, H.~Rasmussen, and J.~Skoustrup.
\newblock Modeling supermarket refrigeration systems for demand-side
  management.
\newblock {\em energies}, 6:900--920, 2013.

\bibitem{GreyBox}
Kenneth Leerbeck, Peder Bacher, and Christian Heerup.
\newblock Grey box modelling of supermarket refrigeration room.
\newblock In {\em Proceedings of the 2021 International Conference on
  Electrical, Computer and Energy Technologies}. European Union, 2021.
\newblock 2021 International Conference on Electrical, Computer and Energy
  Technologies, ICECET 2021 ; Conference date: 09-12-2021 Through 10-12-2021.

\bibitem{madsen2007time}
Henrik Madsen.
\newblock {\em Time series analysis}.
\newblock Chapman and Hall/CRC, 2007.

\bibitem{coolprop}
Ian~H. Bell, Jorrit Wronski, Sylvain Quoilin, and Vincent Lemort.
\newblock Pure and pseudo-pure fluid thermophysical property evaluation and the
  open-source thermophysical property library coolprop.
\newblock {\em Industrial \& Engineering Chemistry Research}, 53(6):2498--2508,
  2014.

\bibitem{TSA}
Henrik Madsen.
\newblock {\em Time Series Analysis}.
\newblock Chapman \& Hall/CRC - Taylor \& Francis Group, 2007.

\bibitem{Fredslund}
K.~Fredslund.
\newblock Load profiles for supermarket refrigeration.
\newblock Technical report, IPU, 2013.

\bibitem{Dormann}
C.~F. Dormann.
\newblock Collinearity: a review of methods to deal with it and a simulation
  study evaluating their performance.
\newblock {\em ECOGRAPHY}, 36:27--46, 2009.

\bibitem{Hair}
J.~F. Hair.
\newblock {\em Multivariate data analysis with readings}.
\newblock Prentice-Hall, 1995.

\end{thebibliography}
